\documentclass[%
 reprint,
superscriptaddress,
 amsmath,amssymb,
 aps,
prd,
longbibliography
]{revtex4-1}

\usepackage{graphicx}
\usepackage{dcolumn}
\usepackage{bm}
\usepackage{hyperref}
\usepackage{color}
\usepackage[mathlines]{lineno}
\usepackage{tabularx}
\usepackage{subfigure}
\usepackage[normalem]{ulem}
\usepackage[export]{adjustbox}
\usepackage{url}
\hypersetup{
    pdfnewwindow=true,    
    colorlinks=true,      
    linkcolor=blue,       
    citecolor=blue,       
    filecolor=blue,       
    urlcolor=blue         
}

\newcommand\chandra{{\it Chandra}}
\newcommand\xmm{{\it XMM-Newton}}

\newcommand\integral{{\it INTEGRAL}}

\newcommand\nustar{\hbox{\it NuSTAR}}
\newcommand\fermi{{\it Fermi}}

\newcolumntype{Y}{>{\centering\arraybackslash}X}

\def\arcmin{\hbox{$^\prime$}}
\def\arcsec{\hbox{$^{\prime\prime}$}}

\usepackage{natbib}

\begin{document}
\title{New~Constraints~on~Sterile~Neutrino~Dark~Matter~from~\textit{NuSTAR}~M31~Observations}

\author{Kenny C. Y. Ng}
\email{chun-yu.ng@weizmann.ac.il}
\thanks{\scriptsize \!\! \href{http://orcid.org/0000-0001-8016-2170}{orcid.org/0000-0001-8016-2170}}
\affiliation{
Department of Particle Physics and Astrophysics, Weizmann Institute of Science, Rehovot, Israel}

\author{Brandon M. Roach}
\email{roachb@mit.edu}
\thanks{\scriptsize \!\! \href{http://orcid.org/0000-0003-3841-6290
}{orcid.org/0000-0003-3841-6290
}}
\affiliation{Department of Physics, Massachusetts Institute of Technology, Cambridge, MA 02139, USA}
\author{Kerstin Perez}
\email{kmperez@mit.edu}
\thanks{\scriptsize \!\! 
\href{https://orcid.org/0000-0002-6404-4737}{orcid.org/0000-0002-6404-4737}}
\affiliation{Department of Physics, Massachusetts Institute of Technology, Cambridge, MA 02139, USA}
\author{John~F.~Beacom}
\email{beacom.7@osu.edu}
\thanks{\scriptsize \!\!  \href{http://orcid.org/0000-0002-0005-2631}{orcid.org/0000-0002-0005-2631}}
\affiliation{Center for Cosmology and AstroParticle Physics (CCAPP), Ohio State University, Columbus, OH 43210, USA}
\affiliation{Department of Physics, Ohio State University, Columbus, OH 43210, USA}
\affiliation{Department of Astronomy, Ohio State University, Columbus, OH 43210, USA} 
\author{Shunsaku Horiuchi}
\email{horiuchi@vt.edu}
\thanks{\scriptsize \!\!  \href{http://orcid.org/0000-0001-6142-6556}{orcid.org/0000-0001-6142-6556}}
\affiliation{Center for Neutrino Physics, Department of Physics, Virginia Tech, Blacksburg, VA 24061, USA}
\author{Roman Krivonos}
\email{krivonos@iki.rssi.ru}
\thanks{\scriptsize \!\!  \href{http://orcid.org/0000-0003-2737-5673}{orcid.org/0000-0003-2737-5673}}
\affiliation{Space Research Institute of the Russian Academy of Sciences (IKI), Moscow, Russia, 117997}
\author{Daniel R. Wik}
\email{wik@astro.utah.edu}
\thanks{\scriptsize \!\!  \href{http://orcid.org/0000-0001-9110-2245}{orcid.org/0000-0001-9110-2245}}
\affiliation{Department of Physics and Astronomy, University of Utah, Salt Lake City, UT 84112, USA}

\date{April 22, 2019}

\begin{abstract}
We use a combined 1.2\,Ms of \nustar{} observations of M31 to search for X-ray lines from sterile neutrino dark matter decay. For the first time in a \nustar{} analysis, we consistently take into account the signal contribution from both the focused and unfocused fields of view. We also reduce the modeling systematic uncertainty by performing spectral fits to each observation individually and statistically combining the results, instead of stacking the spectra. We find no evidence of unknown lines, and thus derive limits on the sterile neutrino parameters. Our results place stringent constraints for dark matter masses $\gtrsim 12$\,keV, which reduces the available parameter space for sterile neutrino dark matter produced via neutrino mixing~(\textit{e.g.}, in the $\nu$MSM) by approximately one-third. Additional \nustar{} observations, together with improved low-energy background modeling, could probe the remaining parameter space in the future. Lastly, we also report model-independent limits on generic dark matter decay rates and annihilation cross sections. 
\end{abstract}


\maketitle

\section{\label{sec:intro} Introduction}
\par Numerous lines of evidence from gravitational signatures point to the existence of beyond the Standard Model matter---dark matter~(DM)---that constitutes more than five times the cosmic energy density of baryons~\cite{Bertone:2004pz, Strigari:2013iaa, Buckley:2017ijx}. The identification of DM is an important task in modern science and could lead to the resolution of many outstanding problems in particle physics and cosmology.

\par A powerful DM search strategy is to look for signatures of DM decaying or annihilating into visible products, \textit{i.e.}, indirect detection.  In particular, channels with monoenergetic photons in the final state are powerful search modes due to the efficient separation of signal and background, as the latter is often dominated by a smooth continuum emission. 

\par Many well-motivated DM candidates could lead to line signatures.  In the X-ray band, one of the most studied candidates is sterile neutrinos~\cite{Kusenko:2009up, Adhikari:2016bei, Abazajian:2017tcc, Boyarsky:2018tvu}, which can radiatively decay into an active neutrino and a monoenergetic photon~($\chi\rightarrow \gamma + \nu$) with energy equal to half of the DM mass~\cite{Shrock:1974nd, Pal:1981rm, Dolgov:2000ew, Abazajian:2001vt}. The production of sterile neutrino DM can be naturally achieved in the early universe via a small mixing with active neutrinos~\cite{Dodelson:1993je}, which may be enhanced by the presence of primordial lepton asymmetry~\cite{Shi:1998km}.
As the mixing angle determines both the abundance and decay rate, there is a finite window in the mass-mixing angle parameter plane in which sterile neutrinos could constitute the full DM abundance, thus allowing this scenario to be fully testable. Closing this window would imply additional physics and production mechanisms are needed to make sterile neutrinos a viable DM candidate~\cite{Shaposhnikov:2006xi, Kusenko:2006rh, Petraki:2007gq, Merle:2013wta,Frigerio:2014ifa, Lello:2014yha, Merle:2015oja, Patwardhan:2015kga}. The existence of sterile neutrino DM could provide strong clues for explaining neutrino mass and baryogenesis~\cite{Fukugita:1986hr}, such as the scenario advocated in the $\nu$MSM model~\cite{Asaka:2005an, Asaka:2006nq, Canetti:2012vf,Canetti:2012kh}.

Due to several sensitive X-ray instruments, such as \textit{Chandra}, \textit{Suzaku}, \textit{XMM-Newton}, and \textit{INTEGRAL}, stringent constraints on X-ray line emission have been obtained using many different observations~(\textit{e.g.}, Refs.~\cite{Watson:2006qb, Loewenstein:2008yi, Yuksel:2007xh, Boyarsky:2007ge, Boyarsky:2007ay, RiemerSorensen:2009jp, Horiuchi:2013noa}). Interest in these topics was heightened with the tentative detection of a 3.5-keV line from cluster observations~\cite{Bulbul:2014sua}, which was followed up by many observational studies~\cite{Boyarsky:2014jta, Riemer-Sorensen:2014yda, Jeltema:2014qfa, Boyarsky:2014ska, Malyshev:2014xqa, Anderson:2014tza, Urban:2014yda, Tamura:2014mta, Sekiya:2015jsa, Figueroa-Feliciano:2015gwa, Riemer-Sorensen:2015kqa, Iakubovskyi:2015dna, Jeltema:2015mee, Ruchayskiy:2015onc, Franse:2016dln, Bulbul:2016yop, Hofmann:2016urz, Neronov:2016wdd, Aharonian:2016gzq, Perez:2016tcq, Cappelluti:2017ywp, Tamura:2018scp}. The nature of this line is still inconclusive. The line could be a signature of sterile neutrino DM~\cite{Abazajian:2014gza} or other candidates~\cite{Finkbeiner:2014sja, Higaki:2014zua, Brdar:2017wgy, Namjoo:2018oyn, Nakayama:2018yvj}. However, as the line flux is weak, astrophysical modeling systematics~\cite{Jeltema:2014qfa, Urban:2014yda} or new astrophysical processes~\cite{Gu:2015gqm, Gu:2017pjy} could also be the explanation.  New detectors~\cite{Nandra:2013shg,Kitayama:2014fda, Figueroa-Feliciano:2015gwa, Tamura:2018scp} or techniques, such as velocity spectroscopy~\cite{Speckhard:2015eva, Powell:2016zbo}, are likely required to fully determine its nature.  (Recently, Ref.~\cite{Dessert:2018qih} claims that blank-sky observations with \textit{XMM-Newton} disfavor the DM interpretation of the 3.5-keV line. On the other hand, Ref.~\cite{Boyarsky:2018ktr} claims detection of the 3.5\,keV line in the Milky Way halo up to $35^{\circ}$ with \textit{XMM-Newton}, and refutes the claim of Ref.~\cite{Dessert:2018qih}.)

For X-ray searches of DM, \nustar{} provides unique capabilities in the hard X-ray band, filling a sensitivity gap that persisted for many years~\cite{Boyarsky:2006fg}~(partially covered by \fermi-GBM~\cite{Ng:2015gfa}).  The first \nustar{} DM search was performed with focused observations of the Bullet Cluster~\cite{Riemer-Sorensen:2015kqa}.  But it was soon realized that by taking advantage of the open telescope design and using ``0-bounce photons"~(photons that enter the detector without passing through the reflecting optics), a larger~(unfocused) field of view~(FOV) can be achieved, thus boosting the sensitivity to the diffuse DM emission. Stringent constraints were then obtained using the extragalactic background~\cite{Neronov:2016wdd} and Galactic Center observations~\cite{Perez:2016tcq}.  
Interestingly, a 3.5-keV line is also included in the \nustar{} instrumental background~\cite{Wik:2014}. The background nature of this line has been questioned in Ref.~\cite{Neronov:2016wdd}. However, Ref.~\cite{Perez:2016tcq} found evidence of the 3.5-keV line in Earth-occulted data, which suggests that the line has a detector background origin. The line is also close to the energy threshold of the detector; detailed studies of its possible instrumental origin are ongoing. Future work from the \nustar{} collaboration is expected to elucidate the nature of this line in the \nustar{} data. 

\par In this work, we search for DM lines with a combined 1.2\,Ms of observations of M31, and offer several improvements on previous works.  Compared to Ref.~\cite{Perez:2016tcq}, the reduction of the astrophysical background, especially from astrophysical emission lines, improves the sensitivity to DM.  We also consistently include both focused (2-bounce photons) and unfocused (0-bounce photons) FOV in signal modeling.  To reduce potential systematic errors from stacking different data sets, we devise a method to statistically combine sensitivities from individual observations.  

\par We present our \nustar{} data analysis in Sec.~\ref{sec:analysis}, our DM signal modeling in Sec.~\ref{sec:dmsignal}, and our DM results in Sec.~\ref{sec:constraints}. We conclude in Sec.~\ref{sec:conclusions}.

\section{\label{sec:analysis} \textit{ NuSTAR } Data Analysis}

In this section, we describe the \nustar{} instrument, the \nustar{} M31 observations, and details about the 0-bounce and 2-bounce FOV.  We comment on the current difficulties in modeling the low-energy \nustar{} background.  Lastly, we detail the modeling of the spectral data.   

\subsection{\label{sec:instrument} The \textit{NuSTAR} instrument}
\par As the first focusing hard X-ray observatory, \nustar{} provides a unique platform for studying astrophysical phenomena---including light DM candidates such as keV-scale sterile neutrinos. The \nustar{} design is detailed in Ref.~\cite{Harrison:2013}, but we re-state some relevant aspects here. 
\par The \nustar{} science instrument consists of two independent, co-aligned focal-plane modules (FPMs), with each FPM consisting of an X-ray optic and a detector. The energy resolution is set by the X-ray detectors, which have a FWHM of 0.4 keV for 5-keV photons, increasing to 0.9 keV for 70-keV photons. The X-ray optics are conical approximations of the grazing-incidence Wolter-I design, with nested Pt/C multilayer-coated mirrors. After entering the telescope, X-rays reflect once off a parabolic mirror segment, followed by a hyperbolic mirror segment. These correctly focused X-rays are called ``2-bounce'' photons, as they reflect twice inside the optics. \nustar{} is sensitive to 2-bounce photons with energies 3--79 keV. The lower limit is due to absorption by the dead layer of the CdZnTe detector, the Pt contact coating, and the 100$\,\mu$m-thick Be window, effects which become significant only for $ E < 10$\,keV~\cite{Madsen:2015}.
The upper limit is set by the Pt K-edge of the optics. Both FPMs have the same essentially overlapping 13\arcmin$\times$13\arcmin~FOV for 2-bounce photons.

\par The \nustar{} optical elements and detectors are separated by a 10-m mast, which is open to the sky. The observatory therefore includes a series of aperture stops to limit unfocused X-rays from striking the detectors. However, this shielding is not complete, and there remains a circular region of radius $\sim3.5^\circ$ on the sky (partially blocked by the optics bench), from which X-rays can strike the detectors without interacting with the optical elements. These are called ``0-bounce'' photons. The energy range for these 0-bounce photons is not constrained by the performance of the optics, and thus extends up to the instrumental limit of $165$ keV.  Similar to the 2-bounce photons, 0-bounce photons are also subject to absorption effects from detector components at low energy.

\subsection{\label{sec:m31obs} \textit{NuSTAR}'s view of M31}
\begin{table*}
\centering
\caption{\nustar{} observations of M31 used in this analysis, with 0-bounce effective areas after data cleaning.}
\begin{tabularx}{\textwidth}{YYYYY}
\hline
\hline
\textit{NuSTAR} obsID & Pointing (J2000) & Effective Exposure\footnote{After \texttt{OPTIMIZED} SAA filtering and solar flare removal.}  & Detector Area $A_\text{0b}$\footnote{After bad pixel removal and point source masking.} & Solid Angle $\Delta \Omega_\text{0b}$\footnote{Average solid angle of sky from which 0-bounce photons can be detected, after correcting for removal of bad pixels, point source masking, and efficiency due to vignetting effects.} \\
 & RA, DEC (deg) & FPMA / B (ks) & FPMA / B (cm$^2$) & FPMA / B (deg$^2$) \\
 \hline
  50026002001 & 11.0826,\: 41.3762  & 95.4 / 94.5 & {11.76 / 11.77} & {4.44 / 4.51} \\
50026002003 &  11.0821,\: 41.3688  & 82.4 / 82.2 & {11.85 / 11.80}  & {4.45 / 4.55}  \\
50026003002 & 11.3306,\: 41.5763  & 106.3 / 105.4 & {11.24 / 11.10} & {4.40 / 4.41} \\
 50110002002 & 11.1122,\: 41.3753 & 31.6 / 31.9 & {12.38 / 12.29} & {4.56 / 4.55} \\
 50110002006 & 11.1047,\: 41.3758 & 36.9 / 36.8 & {12.22 / 12.15}  & {4.55 / 4.52}  \\
   50110003002 & 11.3425,\: 41.5610 & 87.9 / 87.4  & {11.33 / 11.20}  & {4.52 / 4.52} \\
   50111002002 & 11.1285,\: 41.3694 & 82.4 / 82.4  & {11.78 / 11.81}  & {4.46 / 4.51} \\
 50111003002 & 11.3704,\: 41.5913 & 102.0 / 102.2  & {11.22 / 11.16}   & {4.55 / 4.38} \\
 Stacked\footnote{The $A_\text{0b}$ and $\Delta\Omega_\text{0b}$ for the stacked spectra are the exposure-time-weighted averages of the values for the individual observations. } & --- & 624.9 / 622.8 & {11.59 / 11.54} & {4.48 / 4.48} \\
 \hline
 \hline
\end{tabularx}
 \label{obs_table}
\end{table*}

\par 
To probe the diffuse X-ray emission from the direction of M31, we use both \nustar{}'s 0-bounce photons from the wide-angle unfocused FOV and the 2-bounce photons from the narrower focused FOV. Unlike our previous work \cite{Perez:2016tcq}, none of the observations used in the present analysis were contaminated by significant ``stray light'' or ``ghost rays'' resulting from bright, isolated off-axis X-ray sources~(the latter of which are sometimes referred to as ``1-bounce'' photons).

\par Table~\ref{obs_table} shows the eight observations we select from $>1.6$\,Ms of \nustar{} pointed observations of M31 from 2015 to 2018.  These selected observations are optimized to avoid the bright emission from the center of M31 and contamination from bright point sources. 
M31 is the closest large galaxy to our own Milky Way, at a distance $\sim 785$ kpc \cite{Stanek:1998,Holland:1998br}---close enough to resolve bright X-ray point sources. (Specifically, the \nustar{}~18$^{\prime\prime}$ FWHM angular resolution for 2-bounce photons corresponds to $\sim$ 70 pc at the distance of M31.)  We select \nustar{} observations that include at most 1 or 2 resolved X-ray sources in the 2-bounce FOV. We then remove from our analysis all detector pixels corresponding to a radius 60\arcsec~around CXO 004429.57+412135.1 and CXO 004527.34+413253.5, and 100\arcsec~around CXO 004545.57+413941.5 \cite{Lazzarini:2018}. No other point sources with \textit{Chandra} flux (0.35--8.0 keV) greater than $\sim 2\times 10^{-13}$ erg cm$^{-2}$ s$^{-1}$ have been reported in the 2-bounce FOV for any of the observations listed in Table \ref{obs_table}. The remaining M31 emission from faint sources~\cite{Stiele:2018udl} is included in the spectral modeling.

\par Data reduction and analysis are performed with the \textit{NuSTAR} Data Analysis Software pipeline, \textit{NuSTARDAS} \textsc{v1.5.1}. To minimize the charged-particle background due to \nustar{} passing through the South Atlantic Anomaly (SAA), we use the flags \texttt{SAAMODE=OPTIMIZED} and \texttt{TENTACLE=YES}. We also inspect the 3--10 keV light-curves for each observation and remove any time intervals with an elevated low-energy count rate that could be indicative of solar flares. There were three such observations (50026002001, 50026002003, and 50111002002), from each of which 5--10 ks of data were removed. After all the data cleaning, the total exposure time for both FPMs that is used in this analysis is $\sim1.2$ Ms. 

\par
\begin{figure*}
\centering
\begin{minipage}[b]{.48\textwidth}  
    \includegraphics[scale=0.56,right]{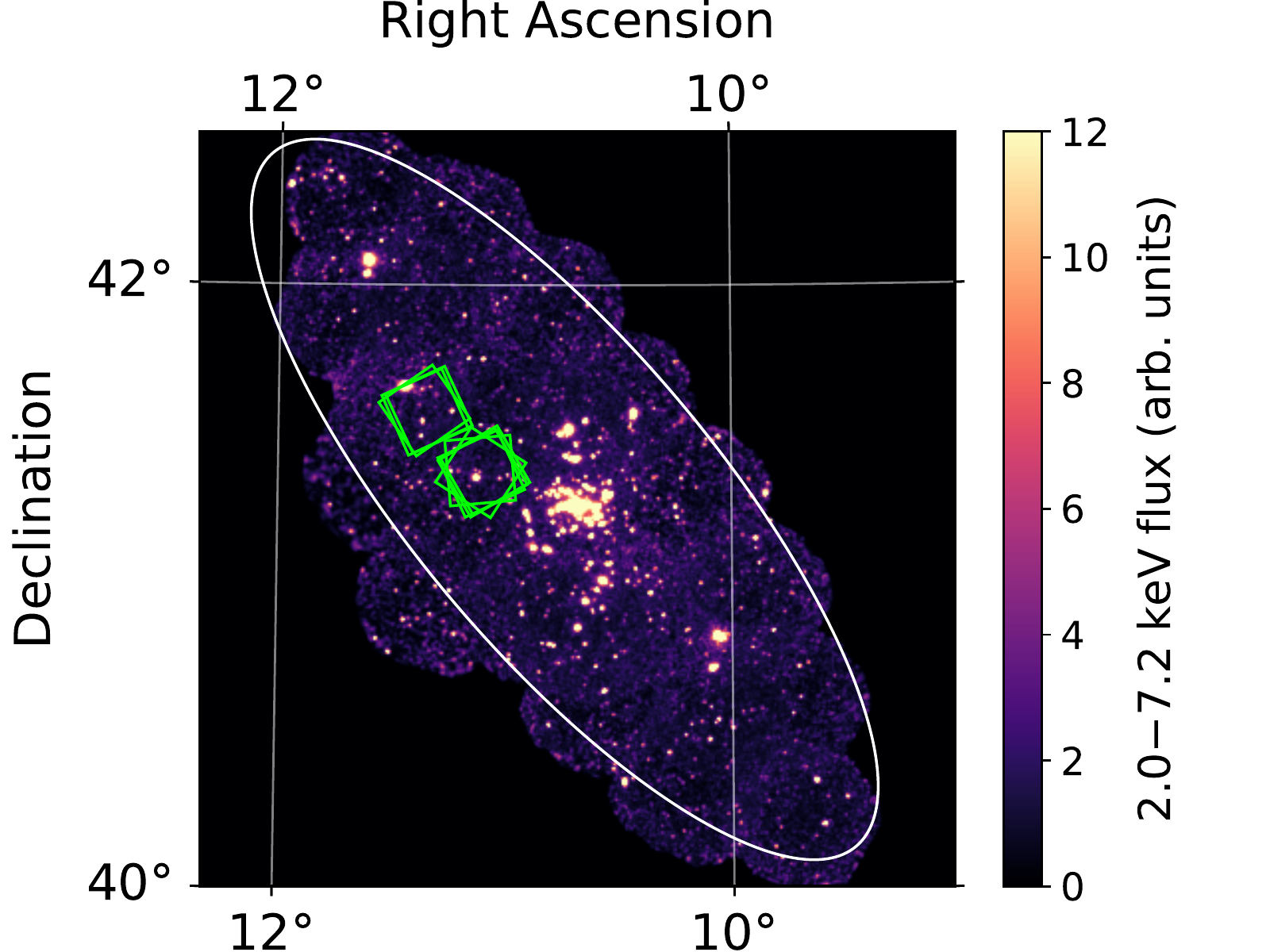}
\end{minipage}\qquad
\begin{minipage}[b]{.48\textwidth}
    \includegraphics[scale=0.56,right]{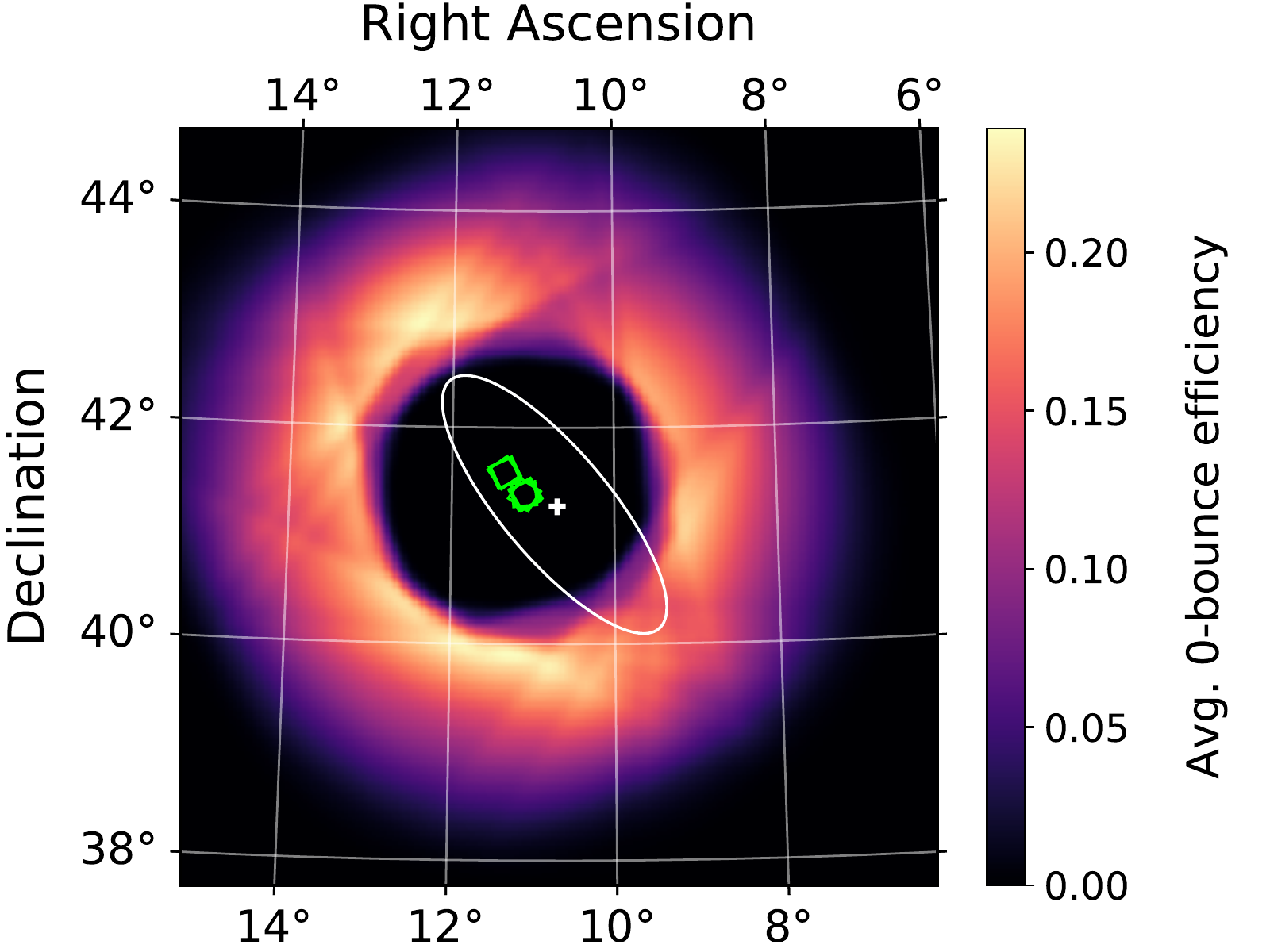}
\end{minipage}
\caption{ \textit{\textbf{Left:}} Magnified view of the M31 disk region. The green squares represent the 2-bounce FOV, overlaid on the
X-ray map by the \xmm{} EPIC instrument \cite{Stiele:2011,xmm_m31_web}, convolved with a Gaussian blur of radius 2 pixels.
The colorbar indicates the 2.0--7.2 keV flux. The brightest X-ray sources in the FOV are removed in the analysis as described in Sec.~\ref{sec:m31obs}. The white ellipse indicates the approximate optical size of the M31 disk.
\textit{\textbf{Right:}} A zoomed-out version of the left, with the color map indicating the averaged efficiency of the 0-bounce FOV for all observations~(FPMA+B) listed in Table~\ref{obs_table}.  The average is weighted by each observation's exposure time after data cleaning and point source removal, and accounting for blocking by the optics bench and vignetting due to the aperture stops. The white cross indicates the center of M31. }
\label{fig:0b_2b_FOV}
\end{figure*}

 Figure~\ref{fig:0b_2b_FOV} shows the the combined 0-bounce and 2-bounce sky coverage of these observations. The 2-bounce FOV avoids the bright astrophysical X-ray emission from the central region of M31, but is still near the center of the DM density distribution. Similarly, the 0-bounce FOV avoids much of the astrophysical X-ray emission from the M31 disk, but is still within the DM halo. For reference, the $\sim 200$\,kpc virial radius of the M31 halo ~\cite{2012A&A...546A...4T} corresponds to $\sim$ $15^\circ$ on the sky.

\subsection{\label{sec:0b2b} Combining 0-bounce and 2-bounce observations}
\par Unlike previous \nustar{} sterile neutrino searches, which considered 0-bounce \cite{Perez:2016tcq, Neronov:2016wdd} and 2-bounce \cite{Riemer-Sorensen:2015kqa} photons separately, the present analysis consistently incorporates both. Although the 2-bounce FOV for these M31 observations is over two orders of magnitude smaller than the corresponding 0-bounce FOV, including the 2-bounce photons increases the sensitivity of our search for two reasons. First, as the DM density increases toward the center of M31, the integrated DM densities over the 2-bounce FOV can be higher than that of the 0-bounce FOV.  Second, the 2-bounce effective area is larger than the 0-bounce effective area, and is maximized for $E \sim10$ keV, which is an especially interesting energy range for sterile-neutrino DM search. 

\par To take advantage of all available data, we extract spectra from the full detector planes. Our spectral model must then account for the instrumental background, as well as astrophysical emission observed in 2-bounce and 0-bounce modes. We account for these multiple spectral components by assigning custom response files for the 0-bounce and 2-bounce spectral model components with the corresponding effective area~(cm$^2$) and effective solid angle~(deg$^2$) factors. 

\par The energy dependent \nustar{} 2-bounce effective area $A_\text{2b}(E)$ is determined primarily by the optical elements, and is calculated by \textit{NuSTARDAS} for each observation. The nominal effective area for each FPM for \textit{point sources} peaks at $\sim 500$\,cm$^2$ at 10\,keV~\cite{Harrison:2013}. For this analysis, the peak effective area is reduced to $\sim 100$\,cm$^2$ due to two reasons: first, the removed point sources are typically near the \textit{NuSTAR} optical axis, where the effective area is the largest; second, the effect of vignetting as the spectra are extracted from the entire FPM as an extended source, where $A_\text{2b}$ is averaged over the 2-bounce FOV. The 2-bounce solid angle $\Delta \Omega_\text{2b}$ is also slightly reduced from $13^\prime \times 13^\prime$ (0.047 deg$^2$) to $\sim 0.045$\,deg$^2$ for each FPM following point source removal. When fitting the 2-bounce components in the spectrum, we use the combined 2-bounce response $ A_\text{2b}(E)\times \Delta \Omega_\text{2b}$, where the $A_\text{2b}(E)$ produced from \textit{NuSTARDAS} already includes the Be window and the detector absorption effects.

\par The effective area $A_\text{0b}$ for 0-bounce photons is set by the physical $\sim$ 15 cm$^2$ area of each detector, and is reduced to $\sim$ 11.5--12.5 cm$^2$ per detector after removing point sources. This is balanced, however, by an increased FOV compared to 2-bounce photons. Using the geometric model of \nustar{} in the \texttt{nuskybgd} code \cite{Wik:2014}, we calculate the average solid angle $\Delta \Omega_\text{0b}$ from which 0-bounce photons can strike the detectors, including the effects of obscuration and vignetting introduced by the optics bench and aperture stop. Following data cleaning and point source removal, each FPM subtends an average solid angle $\Delta \Omega_\text{0b}\sim 4.5$ deg$^2$, almost two orders of magnitude larger than $\Delta \Omega_\text{2b}$. (These parameters are listed in Table~\ref{obs_table}.) For 0-bounce spectral components, we use the combined 0-bounce response $\mathcal{E}_\text{Be}(E)\times A_\text{0b}\times \Delta\Omega_\text{0b}$, with the detector absorption effects included during spectral modeling. Additionally, the use of 0-bounce photons means that we are not limited to the 3--79 keV energy range set by the \nustar{} optics; rather, we can extend our high-energy range up to $E = 100$ keV. Above that, we expect our instrumental background model to be less robust for line searches.

\begin{figure*}[t]
    \centering
    \includegraphics[width=\textwidth]{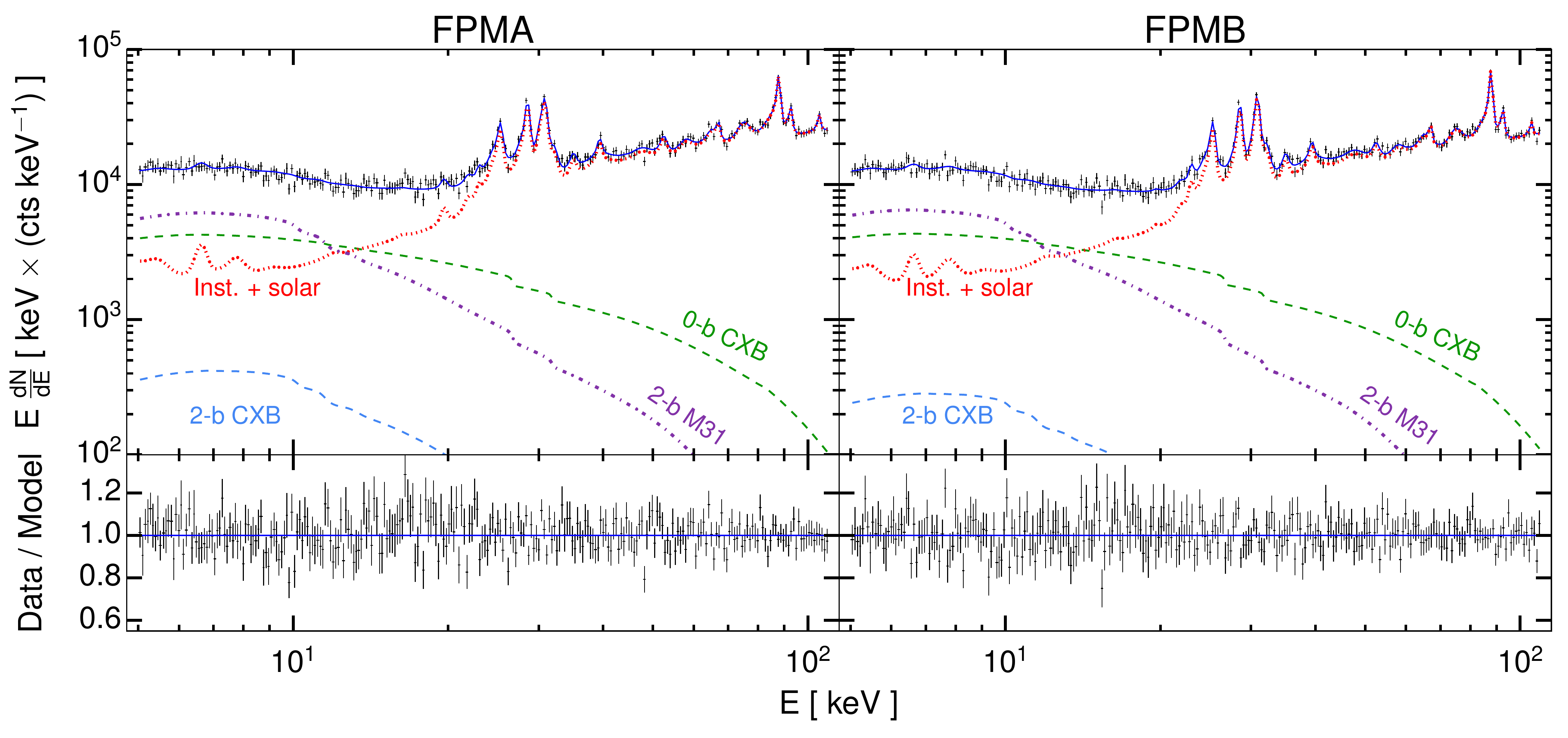}
    \caption{Data and model spectra from FPMA~(left) and FPMB~(right) for the example of obsID 50026002003, including contributions from 0-bounce CXB, 2-bounce CXB, 2-bounce M31, and instrumental/solar background. The 0-bounce M31 component is not included, as the M31 disk is blocked in the 0-bounce FOV.  See Sec.~\ref{sec:spectfit} for details. The lower panels show the ratio of the data to the best-fit model. All error bars indicate $1\sigma$ statistical uncertainties, with reduced $\chi^2$ of 1.15 and 0.99 for FPMA and FPMB, respectively. The differing contributions for the 2-bounce CXB component between FPMA/B arise primarily from differences in the position of the masked point source with respect to the optical axis, as discussed in Sec.~\ref{sec:0b2b}. Results for other observations are similar. }
    \label{fig:spectrum}
\end{figure*}

\subsection{\label{sec:lowenergy} \textit{NuSTAR} at low energy and the 3.5-keV line}
Previous analyses have noted the presence of a line in the \nustar{} spectrum near 3.5 keV~\cite{Neronov:2016wdd,Wik:2014}. These works differ, however, in whether this lines is attributable to an astrophysical~(including DM) or instrumental origin.  
As the \nustar\ instrumental background is poorly understood below 5\,keV, we do not include this energy range in our analysis.
Instead, we comment here upon the difficulties encountered when using this low-energy \nustar\ data and the foreseen avenues for future progress. 

We investigate our instrumental background components using occulted data, collected during the same M31 observation periods we use for our main analysis but when the \nustar\ FOV is blocked by the Earth. We consider two spectral models for this data: the default \nustar{} instrumental background model and an ``internal power-law" version of this model; the latter is motivated by fits to the occulted data themselves.

The default \nustar\ instrumental background model, as detailed in Sec.~\ref{sec:spectfit} and Ref.~\citep{Wik:2014}, is derived from phenomenological fits to ``blank sky" observations. 
It is dominated at low energies by a 3.5-keV line,  a 4.5-keV line, and an $\sim1$-keV thermal plasma component (the {\verb=apec=} model in {\verb=XSPEC=}) that is possibly attributed to reflected solar X-rays. 
At high energies, it is dominated by a relatively flat continuum and a series of Lorentzian lines. 
We find that for energies above $\sim20$\,keV, the occulted data is well described by this model.
Below $\sim20$\,keV, however, the occulted data indicates residual emission remains that is not accounted for by this default model. 

These low-energy occulted fits are improved if we modify the default model by replacing the $\sim1$\,keV thermal plasma component with a power-law continuum. 
In this ``internal power-law" model, we use occulted data to derive a best-fit power-law index and relative normalization with respect to the high-energy continuum; we then freeze both of these parameters in the instrumental background model applied to non-occulted data. 
This procedure has been validated on extragalactic observations, where it yields the correct expected spectral shape and flux for the Cosmic X-ray Background (CXB)~\cite{Gruber:1999,Churazov:2006bk}.

To associate any detected line with an astrophysical, as opposed to instrumental, origin, the observed line flux should be smaller or nonexistent in the occulted dataset. 
Using the default instrumental background model, the 3.5-keV and 4.5-keV lines are each observed with comparable 90\% C.L. line fluxes between occulted and non-occulted data. 
When we adopt the internal power-law background model to M31 data, the 3.5-keV and 4.5-keV line fluxes are again each observed with comparable 90\% C.L. line fluxes between occulted and non-occulted data. 
However, the best-fit line fluxes differ between the default and power-law instrumental background models, between different observations of similar sky regions, and between FPMA and FPMB of the same observation.  In addition, the fit is unstable when using the internal power-law model below 5\,keV, where the 3.5-keV and 4.5-keV line strengths are found to be somewhat degenerate with the power-law index and normalization. 
This study reinforces the notion that these lines are backgrounds, though the statistics of the occulted sample is relatively limited. As a result, further interpretation, such as searching for any possible excess in low-energy line flux in the non-occulted M31 data, is difficult. 

Due to these issues, we limit our analysis to $E \geq 5$\,keV, or sterile neutrino DM mass $\geq 10$\,keV. 
This allows for a stable spectral fit and robust determination of line-flux limits. 
We see no difference in the $E\geq 5$\,keV results between using the default or power-law internal background model. We use the default background model for the rest of the paper. We further note that the choice of the background model does not affect the $E < 5$\,keV limits derived from the previous \nustar\ Galactic Center analysis \cite{Perez:2016tcq}; this constraint was conservative, allowing the DM line to assume the full strength of any lines. 

Detailed investigations of the \nustar{} instrumental background are beyond the scope of this paper. Work is ongoing now to exploit the full \nustar\ archival data set to better constrain the origin and description of the instrumental background. Future \nustar\ analyses will be able to use this improved model to better constrain, or detect, low-energy line emission.

\subsection{\label{sec:spectfit} Spectral fit}
\par We consider the \nustar{} data between 5--100\,keV, as discussed above. 
We first fit each observation individually with their own set of parameters. In Sec.~\ref{sec:combined_analysis} below, we combine the fits to derive our primary results.  
Our spectral model consists of four components: the default \nustar{} instrumental background~\cite{Wik:2014}, 0-bounce and 2-bounce CXB components, and a 2-bounce component from the diffuse M31 emission. The first component does not depend significantly on the FOV of the observations, while the the rest do. Note that we use both 0-bounce and 2-bounce astrophysical components, normalized as described in Sec.~\ref{sec:0b2b}.

\par Because the number of photons is large, we are able to choose a binning scheme that is sufficiently fine to identify any narrow spectral features, while also minimizing the statistical error in each bin. We adopt a logarithmic binning scheme of 100 bins per decade in energy for the $\sim30$ ks observations (50110002002 and 50110002006), and 200 bins per decade for the remaining, longer observations. These binning schemes provide a statistical uncertainty that is everywhere $\sim 10\%$ per bin. We note that the binning is always narrower than the \nustar{} energy resolution~(FWHM) for photon energies 5--20 keV, which is the most interesting energy range for sterile neutrino DM searches.

\par  We adopt the default \nustar{} background model of Wik \textit{et al.}~\cite{Wik:2014}, including internal detector backgrounds, a solar component, and the CXB. The \nustar{} internal background is modeled by a broken power-law continuum with a break at 124 keV, as well as a complex of  Lorentzian activation and fluorescence lines, which together comprise much of the background above 20 keV. The continuum power-law index, as well as the line energies and widths, are fixed, while the normalisations of these components are free to fit independently for all observations and for FPMA and FPMB. This is due to different detector backgrounds and sky regions between observations and FPMs. 
As described in Sec.~\ref{sec:lowenergy}, we also include a $\sim$ 1 keV diffuse thermal plasma component with emission lines, believed to result from solar activity. This is the source of the lines near 6.5 keV and 8 keV. The CXB, resulting from unresolved extragalactic emission, is described by a cut-off power-law with spectral indices, cut-off energy, and flux fixed to the values measured by \integral{} \cite{Churazov:2006bk}.

\par Observations from \textit{ROSAT}~\cite{West:1997}, \xmm{}, and \chandra{}~\cite{Bogdan:2010} reveal a diffuse X-ray component within the disk of M31, thought to result from a population of unresolved point sources. We adopt a single power-law model for the energy spectrum of 2-bounce photons from these unresolved M31 sources, with normalization and spectral index free to fit. We obtain 2-bounce M31 spectral indices in the range $\sim$~1.3--1.7 for the observations in this analysis, consistent with the population of faint X-ray sources in, \textit{e.g.}, Ref.~\cite{Stiele:2018udl}. Since the \nustar{} optics bench blocks the disk of M31, as shown in Fig.~\ref{fig:0b_2b_FOV}, there is no need to include a diffuse 0-bounce M31 component. There is also evidence for at least one thermal plasma component with 0.1~keV $\lesssim kT \lesssim 0.7$ keV to the diffuse M31 X-ray background \cite{Trudolyubov:2004tk}, though such a low-temperature component would not be visible to \nustar{} in the energy range of this analysis. We stress that a conclusive identification of the diffuse M31 X-ray background is not necessary for the present analysis; rather, it is the use of a physically-motivated model that provides a good fit to the data that is most important. 

\par  All non-instrumental components, including astrophysical, DM, and the solar background components, are subject to absorption by materials on the surface of the \nustar{} detectors. This includes absorption from the $\sim$0.11-$\mu$m Pt contact coatings, as well as the $\sim$0.27-$\mu$m layer of inactive CdZnTe on the detector surfaces. Combined, these components cause a $\sim25$\% absorption for 5-keV photons, decreasing to $\sim5$\% for 10-keV photons. (We note the effect of the Be shield is already included in the effective area.) 

\par The astrophysical components are also subject to absorption from the interstellar medium. For the 2-bounce CXB and 2-bounce M31, the absorption is calculated using the \texttt{tbabs} model in \texttt{XSPEC}, which incorporates the abundances given in Ref.~\cite{Anders:1989zg} and the photoionization and absorption cross-sections given in Refs.~\cite{BalucinskiaChurch:1992ia,Yan:1998}. The equivalent neutral hydrogen column density $N_\text{H}$ toward the disk of M31, near the 2-bounce FOV of the observations in this analysis, is $\sim5\times10^{21}$ cm$^{-2}$ \cite{Braun:2009}. The corresponding optical depth is $\tau\sim 0.03$ for 1-keV photons \cite{Predehl1995}, and decreases with energy.  The $N_\text{H}$ value for 2-bounce observations is fixed during spectral modeling.

\par Similarly, for the 0-bounce CXB, we apply a fixed $N_\text{H}$ value of $\sim7\times10^{20}$ cm$^{-2}$, as observed in the direction of M31 without the disk~\cite{Dickey:1990mf,Kalberla:2005ts}. As the 0-bounce $N_\text{H}$ value is almost an order of magnitude lower than the 2-bounce $N_\text{H}$ value, absorption is negligible; we include it for completeness. 

\par For DM lines, as most of the signal comes the 0-bounce observations~(see Sec.~\ref{sec:dmsignal}), only absorption from detector materials is relevant. This is at most a $\sim25$\% effect at 5\,keV; we include it to be conservative and consistent.

\par Figure~\ref{fig:spectrum} shows the spectrum for obsID 50026002003 and the corresponding spectral fit components, as an example. The vertical-axis units on this figure reflect proportional differences in the number of counts $N$ in each bin, to wit:
\begin{equation*}
    E \frac{dN}{dE} = \frac{dN}{d(\ln E)} \simeq \frac{dN}{d(2.3\log_{10}E)}.
\end{equation*}
As the bins are evenly-spaced in $\log_{10}E$, with 200 bins per decade in the example spectra, each bin has a width 0.005 in $\log_{10} E$. The number of counts in the bin with $E \sim 10$ keV, for example, is
\begin{equation*}
    1.2\times 10^4 \text{\;keV}\times(\text{counts\;keV}^{-1})\times \frac{2.3}{200} \approx 130\text{\;counts},
\end{equation*}
corresponding to a $\sim$ 10\% statistical uncertainty per bin. The fit has
reduced $\chi^{2}$ of 1.15 and 0.99 over 239 degrees of freedom for FPMA and FPMB, respectively. We find similar results for all other observations. Thus, we conclude that the fits are acceptable for the individual observations. In Sec.~\ref{sec:constraints}, we discuss details of the fit residuals as well as the procedure for combining observations. 
While the astrophysical components dominate below $\sim 13$\,keV, both the astrophysical continuum and line emission are much lower compared to our previous Galactic Center analysis~\cite{Perez:2016tcq}. This enhances the DM sensitivity, particularly in the mass range $\sim$12--20 keV, where previously the search was limited by strong Fe line emission from the Galactic Ridge X-ray Excess~(GRXE). 

\section{\label{sec:dmsignal} Dark Matter Signal Modeling}

\par In this section, we describe the modeling of DM event rates in \nustar{} observations. 

\subsection{DM Distributions}\label{sec:densityprofiles}
\par To search for DM lines in the \nustar{} observations, we need to compute the expected DM event rate taking into account both the 0-bounce and 2-bounce FOV. Also, DM from the Milky Way~(MW) and M31 halo can both appreciably contribute to the signal. 

\par The expected photon line intensity~(differential flux per solid angle) from sterile neutrino DM decay is
\begin{equation}\label{eq:intensity}
    {\cal I}\equiv \frac{dF}{dEd\Omega} = \frac{\Gamma}{4 \pi m_{\chi}}\int d\ell \left(\rho_{\rm MW} + \rho_{\rm M31}\right) \frac{dN}{dE} \, ,
\end{equation}
where $\Gamma$ is the DM decay rate, $m_{\chi}$ is the DM mass, $\rho$ is the DM density distribution for MW or M31, $\ell$ is the line-of-sight distance that we integrate over, and $dN/dE = \delta(E-m_{\chi}/2)$ is the sterile neutrino DM decay spectrum with the line energy being half of the DM mass. The line width is narrow compared to the \nustar{} energy resolution, thus a delta function approximation is appropriate. 

\par To evaluate the line-of-sight integrals for MW we use the sNFW profile from Ref.~\cite{Perez:2016tcq}, which was motivated by simulations and MW kinematic data~\cite{Calore:2015oya}~(see Ref.~\cite{Perez:2016tcq} for details). For the M31 halo, we use the NFW profile from Ref.~\cite{2012A&A...546A...4T}, where the mass distribution was inferred from a multitude of imaging and kinematic data.  The scale density, scale radius, and the virial radius are $\rho_{s} = 0.418\,{\rm GeV/cm^{3}}$, $r_{s} = 16.5$\,kpc, and $R_{200} = 207$\,kpc, respectively. In the next section, we discuss the impact of the choice of density profiles to the DM sensitivity. 

\par We neglect the cosmological decaying DM contribution to the extragalactic background. The spectrum of this signal is broadened by cosmological redshifts, and the flux is negligible compared to the other components that we consider~\cite{Perez:2016tcq}.

\subsection{DM event rate in \textit{NuSTAR}}\label{sec:evtrate}

The number of DM signal photons, $N_{\rm DM}$, that would be detected by \nustar{} is the DM intensity~(Eq.~(\ref{eq:intensity})), integrated over the line spectrum and convolved with the detector response for that particular observation, 
\begin{eqnarray}\label{eq:evt_rate}
    N_{\rm DM} &=& \sum_{j = \text{0b,2b}} A_{j}T_{\rm obs}\int_{\rm FOV} {\xi}_{j}\,{\cal I}\,d\Omega \\
    &=& \frac{\Gamma }{4 \pi m_{\chi}}T_{\rm obs} \left( A_\text{0b}\Delta\Omega_\text{0b}{\cal J}_\text{0b} +A_\text{2b}\Delta\Omega_\text{2b}{\cal J}_\text{2b} \right) \nonumber \, ,
\end{eqnarray}
where $A_\text{0b,2b}$ are the effective areas for 0-bounce and 2-bounce observations, $T_{\rm obs}$ is the observing time, and $\xi_\text{0b,2b}$ are the pixel efficiencies of the FOV that takes into account, \textit{e.g.}, the optics bench blocking, and $\Delta\Omega = \int_{\rm FOV}{\xi}d\Omega$ is the effective FOV.  We also define the FOV-dependent J-factor,
\begin{equation}
    {\cal J} = \frac{1}{\Delta\Omega}\int d\Omega \,{\xi} \int d\ell  \left(\rho_{\rm MW} + \rho_{\rm M31}\right)  \,
\end{equation}
for both 0-bounce and 2-bounce observations. 

For the 0-bounce FOV, as shown in Fig.~\ref{fig:0b_2b_FOV}, there is not much variation in ${\cal J}_{0b}$ among different observations, given that they basically all point in the same direction. The MW halo also has a larger contribution to the J-factor compared to the M31 halo.  This is because the bulk of M31 center is blocked by the mirror module, so only the outskirts of the M31 contribute to the J-factor. Roughly, the MW part of the ${\cal J}_{0b}$ is about 1.5 to 2.3 times larger than that of M31. For the 2-bounce FOV, the situation is reversed, as they point closer to the center of M31.  Roughly, the M31 part of ${\cal J}_{2b}$ is about 2.1 to 3.4 times larger than that of the MW.

Unlike previous works, where either the 0-bounce~\cite{Riemer-Sorensen:2015kqa} or the 2-bounce contribution are neglected~\cite{Neronov:2016wdd, Perez:2016tcq}, we include both in this work. To see how much the 2-bounce FOV affects the result, we consider the ``enhancement factor,'' 
\begin{equation}
    \left( 1 + \frac{ A_\text{2b}(E) \Delta\Omega_\text{2b}{\cal J}_\text{2b} }{ A_\text{0b}\Delta\Omega_\text{0b}{\cal J}_\text{0b}} \right )\, ,
\end{equation}
which represents the signal enhancement due to the 2-bounce FOV. The enhancement factor is energy dependent due to the 2-bounce effective area $A_\text{2b}(E)$. 

\begin{figure}[t]
    \centering
    \includegraphics[width = \columnwidth]{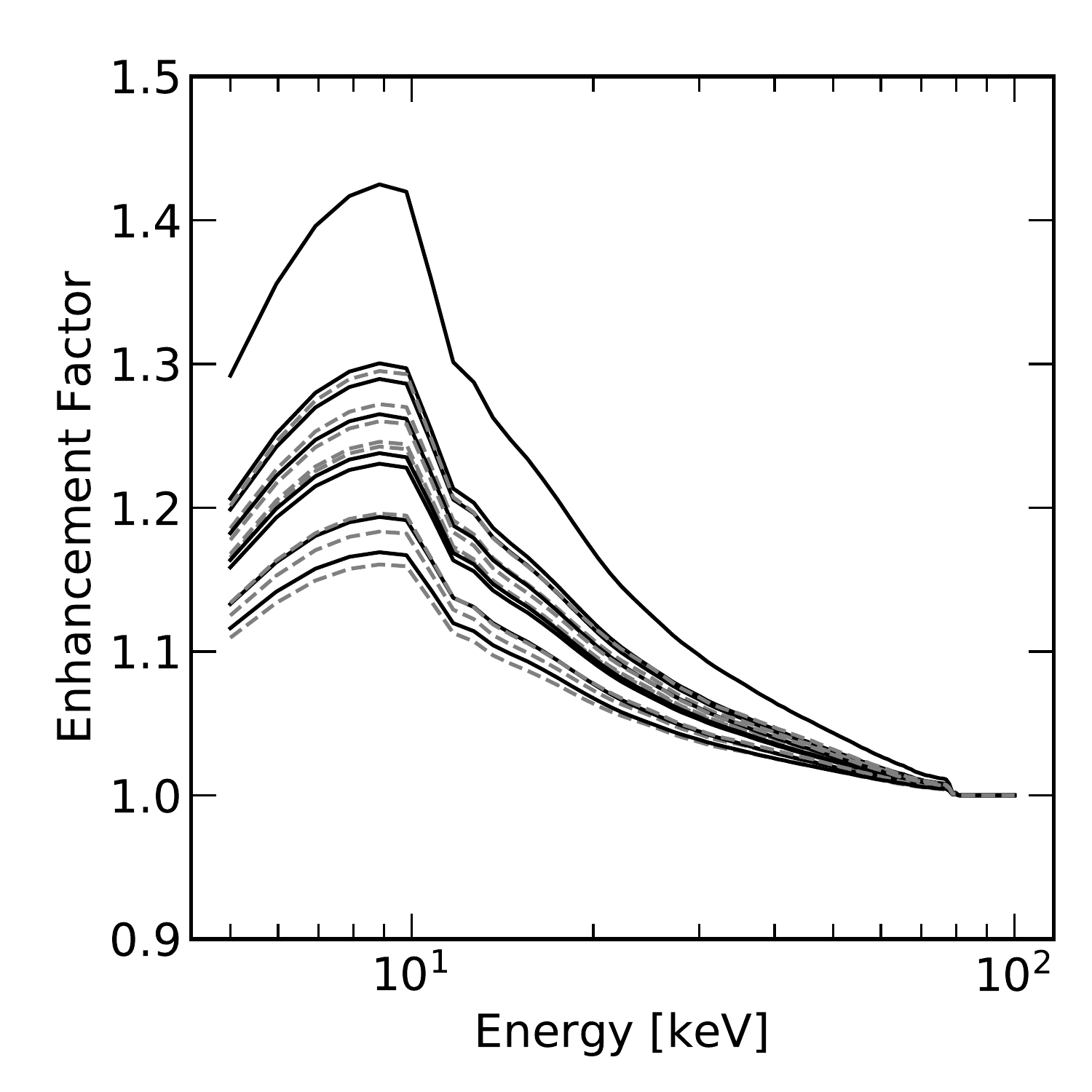}
    \caption{Enhancement due to the inclusion of the 2-bounce FOV relative to the 0-bounce FOV for all observations across FPMA~(solid) and FPMB~(dashed) considered in this work. The energy dependence comes from the effective areas of the 2-bounce FOV.  }
    \label{fig:e_factor}
\end{figure}

Figure~\ref{fig:e_factor} shows the enhancement factors for all the considered observations.  They peak around 10\,keV and can be as high as 1.4. The enhancement from the 2-bounce contribution is negatively impacted by the point source removal and vignetting effects as described in Sec.~\ref{sec:0b2b}, which significantly reduced the effective area. 

After combining the observations~(see Sec.~\ref{sec:combined_analysis}), at $\sim10$\,keV photon energy, the 0-bounce MW, 0-bounce M31, 2-bounce MW, and 2-bounce M31 component contribute about 50\%, 30\%, 5\%, and 15\% of the signal, respectively.

\section{\label{sec:constraints} Dark Matter Line Search}
\par In this section, we describe the line analysis for each individual M31 observation, then the combined analysis. Finally, we present our results for sterile neutrino DM.  

\subsection{ NuSTAR observation line analysis}\label{sec:lineanalysis}
\par For each M31 observation and FPM, we search for potential DM line signals by scanning in the 5--100\,keV energy range, with 100 logarithmically spaced steps per decade in energy. In this scheme, near $E = 5$ keV, the steps are $\sim 0.1$ keV apart, growing to $\sim 2$ keV apart by $E = 100$ keV. This is to be compared to the detector energy resolution of 0.4 keV for 5-keV photons and 0.9 keV for 60-keV photons. Thus, any new line present with $E \lesssim 40$ keV will be several steps wide. We observe no significant change in results if we use a finer sampling scheme, \textit{e.g.}, 200 steps per decade.  For each scanned line energy, we add a hypothetical DM line component to the model and take into account all detector response and absorption effects~(see Sec.~\ref{sec:0b2b}). As the line width is dominated by the detector resolution, the DM line component has only one free parameter---the normalization---for each scan, which we parametrize with the DM decay rate, $\Gamma$, using Eq.~(\ref{eq:evt_rate}).

\par We define the function $\chi^{2}(\Gamma)$, which is the best-fit chi-squared value of the fit to the data after adding the DM signal line to the model at a specific value of $\Gamma$. We find the $\chi^{2}(\Gamma)$ distribution for each scanned line energy. It is important to note that for every value of $\Gamma$, $\chi^{2}(\Gamma)$ is minimized with respect to all free spectral model parameters. Thus, when scanning at a known background line~(detector or astrophysical) energy, $\chi^{2}(\Gamma)$ would be at the minimum value for $\Gamma$ smaller or equal to the corresponding background line strength. In other words, when setting the limit at the presence of a background line, we conservatively allow the DM line flux to be at least as large as the background line feature. 

\subsection{Combined analysis with all M31 observations} \label{sec:combined_analysis}

Having obtained the $\chi^{2}(\Gamma)$ distributions for all observations, we now combine them to take advantage of the full statistical power of the 1.2\,Ms of \nustar{} M31 data. We consider the object $X^{2}$, which is the sum of all the individual $\chi^{2}$ distributions,
\begin{equation}
X^{2}(\Gamma) = \sum_\text{obs} \chi^{2}(\Gamma)\, .   
\end{equation}
In the case of a detection or in the presence of a background line feature, $X^{2}(\Gamma)$ would reach a minimum at a particular $\Gamma_{0}$ value, with the line significance given by $\sqrt{  X^{2}(\Gamma = 0) - X^{2}(\Gamma_{0})}$. In the case of a null detection, the minimum would be at $\Gamma_{0} = 0$.  In all cases, we can obtain the 95\% one-sided upper limit by finding $\Gamma_{95}$ such that $X^{2}(\Gamma_{95}) = X^{2}(\Gamma_{0}) + 2.71$. 

Given that individual observations are minimized independently, they are allowed to have their own set of background parameters.  This is an improvement over previous stacked analyses~\cite{Perez:2016tcq, Neronov:2016wdd}, which used a single background model to describe the sum of many observations. Our new approach avoids producing artificial spectral features due to the stacking of potentially different continuum backgrounds in each observation. In fact, it is known that the background parameters are slightly different between the two FPMs. The new approach also effectively increases the number of d.o.f., which alleviates the need for adding an extra systematic error factor to improve the fit, as in our previous Galactic Center analysis~\cite{Perez:2016tcq}. In the future, this approach can even be used to combine the sensitivities from observations of different targets.  In Appendix~\ref{sec:stacked}, we detail the precedures of the stacked analysis. The results are found to be consistent with our default analysis. 

\begin{figure}[t]
    \includegraphics[width = \columnwidth]{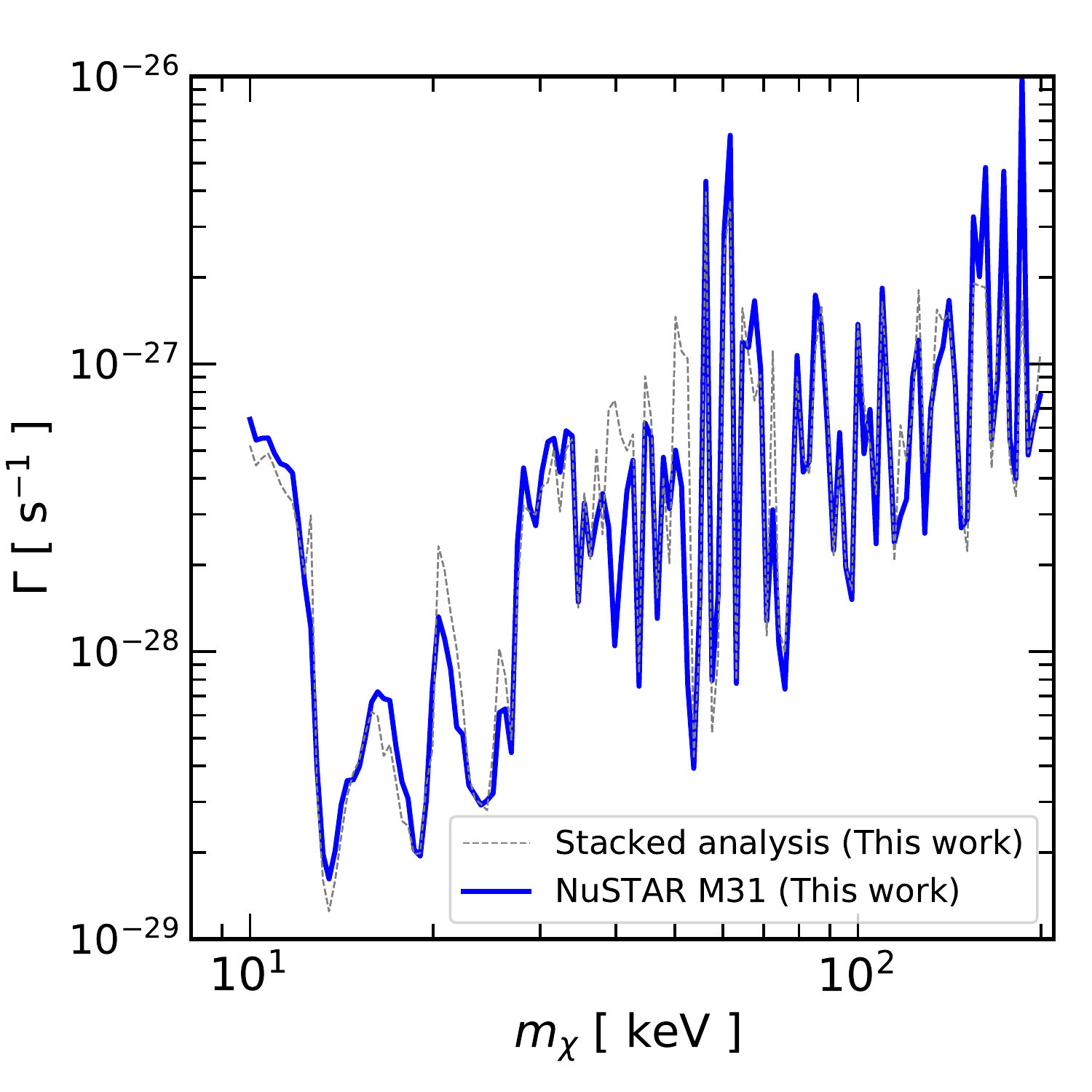}
    \caption{ Derived upper limit on DM decay rates from combined \nustar{} M31 observations. Results from the default analysis are shown with a blue line, and that from the stacked analysis~(see Appendix~\ref{sec:stacked}) are shown witha  grey dashed line.  We have assumed each DM decay produces one monoenergetic photon with  energy half the DM mass.    }
    \label{fig:gamma}
\end{figure}

\begin{figure}[t]
    \includegraphics[width = \columnwidth]{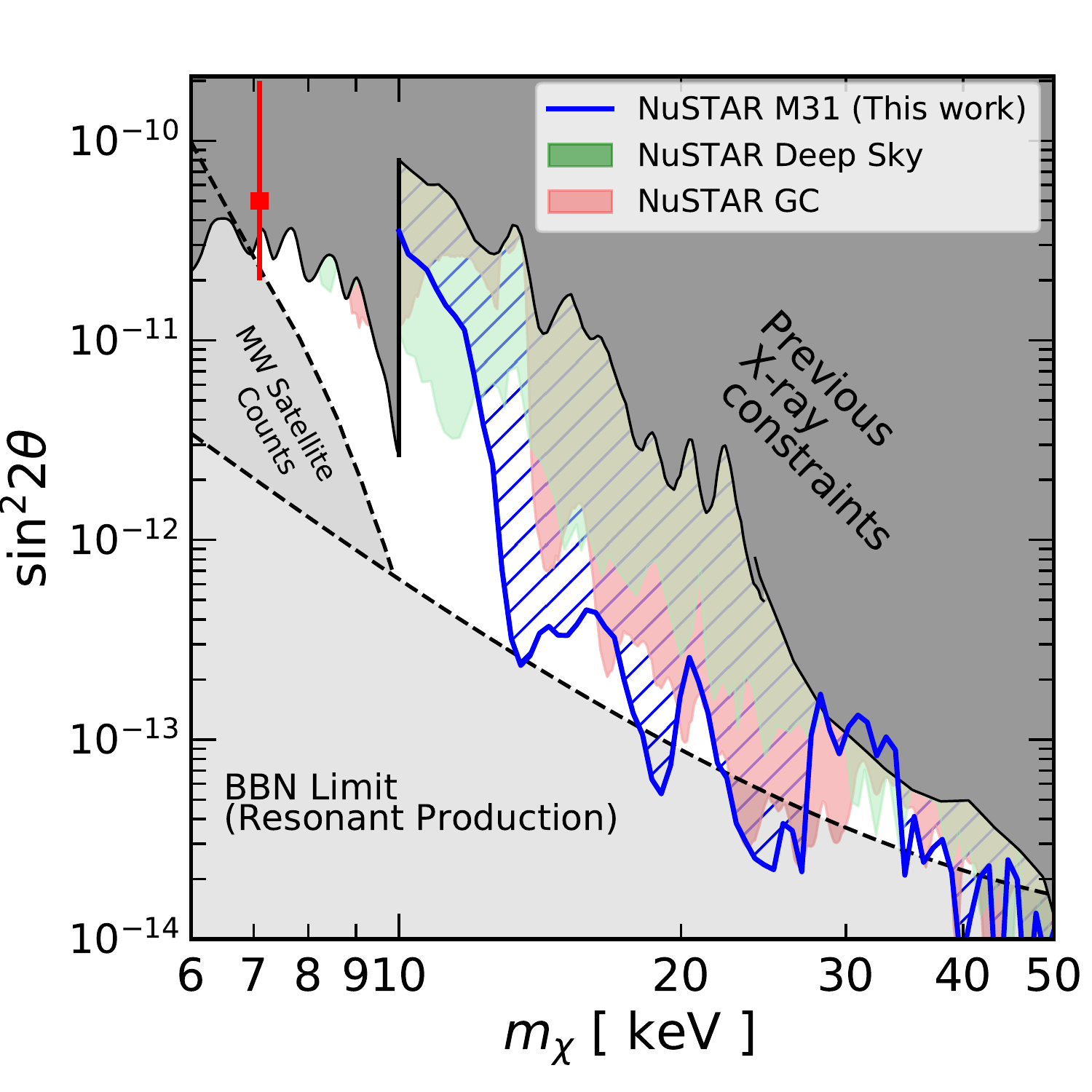}
    \caption{ The mixing angle-mass parameter space for sterile neutrino DM. Our limit obtained from the combined M31 observations is shown by the blue line and the hatched region.  For comparison, we also show \nustar{} constraints from deep sky~\cite{Neronov:2016wdd} and Galactic Center~\cite{Perez:2016tcq} observations. The previous X-ray constraints are shown in the dark grey region~(see Ref.~\cite{Perez:2016tcq} for details).  For sterile neutrino DM produced via mixing, the light grey constraints from satellite counts~\cite{Cherry:2017dwu} and BBN constraints on lepton asymmetry also apply.  In this case, a finite allowed window remains, shown in white. The red point indicates the claimed 3.5\,keV line detection.}
    \label{fig:sndm_constraint}
\end{figure}

In the line search, we identify several energy regions where the fit noticeably improves when a DM line is added. These are at photon energies around 5--6\,keV, 15--16\,keV, 28\,keV, and 80\,keV.  For the 5--6\,keV region, the residual is located at the edge of our selected data. As mentioned in Sec.~\ref{sec:lowenergy}, the low-energy \nustar{} background model is quite uncertain.  In particular, it is likely that the actual solar component differs from that in the default background model, due to the Sun being a variable source. In contrast to our previous Galactic Center analysis~\cite{Perez:2016tcq}, this problem in the instrumental background is highlighted by the low astrophysical flux in the data. 
For 15--16\,keV, it is known that the current background model has difficulty in fitting the data here, likely due to this being the transition region between the astrophysical and instrumental continuum backgrounds. A similar residual was also identified in our previous Galactic Center analysis~\cite{Perez:2016tcq}. 
For the 28\,keV and 80\,keV regions, these are where the data is dominated by strong and complicated detector activation and fluorescence lines, as shown in Fig.~\ref{fig:spectrum}, thus making the astrophysical interpretation difficult and unlikely. We also note that the corresponding line fluxes at 28\,keV and 80\,keV are already being constrainted by \emph{INTEGRAL}~\cite{Yuksel:2007xh,Boyarsky:2007ge}. As a result, we do not interpret all these residuals as having an astrophysical origin.  The corresponding DM limit is set conservatively, with the DM flux allowed to saturate the full residual, thus resulting in worsened sensitivities in these regions.

Overall, we see no obvious candidates of a DM signal. We thus proceed to derive upper limits for the DM decay rate, following the procedure described above. 

Figure~\ref{fig:gamma} shows the derived constraints on the decay rate, $\Gamma$. In the energies mentioned above, where we see positive residuals, our limit is weakened.  In other energies, where there are background lines, our limit is also weakened, causing many ``spikes'' in the limit. For comparison, we also show the results from the stacked analysis, which are consistent with the default analysis. 

Our results are robust with respect to the choice of DM density profiles.  Because M31 is located relatively far away from the Galactic Center in celestial coordinates~($\sim120^{\circ}$), the MW part of the signal is insensitive to the choice of the MW profiles. Using different profiles considered in Ref.~\cite{Perez:2016tcq} only changes the expected MW signal by $\sim10$\%, in contrast to the potential large uncertainties when using the Galactic Center~\cite{Yunis:2018eux}. For the M31 part, because the signal is dominated by the 0-bounce observations, which are only sensitive to the outskirts of the M31 halo~($\gtrsim 10$\,kpc), using different halo profiles considered in Ref.~\cite{Tamura:2014mta} changes the total sensitivity by only $\lesssim5$\%.

While we focus on sterile neutrino DM, the limit on the sterile neutrino DM decay rate can be easily translated and applied to other decaying DM models. In Appendix~\ref{sec:annihilation}, we also derive constraints on the cross section for annihilating DM models.

\subsection{Constraints on Sterile Neutrino DM}

\par We now consider the implications of our result for sterile neutrino DM, one of the prime candidates for decaying DM in the X-ray band.  

\subsubsection{X-ray constraints}

\par Sterile neutrino DM, regardless of the production mechanism, decay radiatively~($\chi\rightarrow\gamma+\nu$) via the mixing with active neutrinos, with a rate~\cite{Shrock:1974nd, Pal:1981rm} 
\begin{equation}\label{eq:decay_rate}
    \Gamma = 1.38\times 10^{-32}\,{\rm s^{-1}}\left( \frac{\sin^{2}2\theta}{10^{-10}} \right)
    \left( \frac{m_{\chi}}{\rm keV} \right)^{5} \, ,
\end{equation}
where $\theta$ is the mixing angle with active neutrinos. 
Thus, X-ray observations can be used to place model-independent upper limits on the mixing angle~\cite{Dolgov:2000ew, Abazajian:2001vt}, which is the main goal of this work. 

Figure~\ref{fig:sndm_constraint} shows the mass-mixing angle parameter space plane. Our constraint from the combined M31 observations, converted from decay rates using Eq.~(\ref{eq:decay_rate}), is shown by the blue line. For comparison, we also show constraints from many previous X-ray searches detailed in Ref.~\cite{Perez:2016tcq}. Compared to our previous Galactic Center analysis~\cite{Perez:2016tcq}, our present results benefit from the absence of strong astrophysical line emission around $6.7$\,keV. Compared to the deep sky analysis~\cite{Neronov:2016wdd}, our present analysis, despite having less exposure, benefits from the inclusion of the DM contribution from the M31 halo, as well as from an improved analysis procedure that minimizes systematic effects from stacking spectra. Overall, our new results improve the constraints for DM masses between 12--20\,keV.

\subsubsection{ A finite window in the parameter space }

If sterile neutrino DM is produced via mixing in the early universe, then additional constraints from DM production and warm DM considerations also apply.  These constraints, together with the X-ray constraints, form a window in the parameter space that is bounded on all sides.  

\par Sterile neutrino DM can be naturally produced via neutrino mixing with active neutrinos, either non-resonantly~\cite{Dodelson:1993je} or resonantly~\cite{Shi:1998km}. Non-resonant production defines a line in the parameter space of mass and mixing angle, and is already in strong tension with exiting constraints. Resonant production can occur if there was primordial lepton asymmetry, which allows sufficient DM to be produced with a range of smaller mixing angles than non-resonant production.  However, Big Bang Nucleosynthesis~(BBN) places an upper limit on the lepton asymmetry at that epoch~\cite{Dolgov:2002ab, Serpico:2005bc, Boyarsky:2009ix}; the limit on the asymmetry is often expressed through the dimensionless parameter $L_{6} \leq 2500$. Using the latest sterile neutrino DM production code, \texttt{sterile-dm}~\cite{Venumadhav:2015pla}, we find the corresponding \emph{lower} bound on the mixing angle; below that neutrino mixing is unable to produce enough DM to match the observed abundance. We note that our result is consistent with that found in Ref.~\cite{Cherry:2017dwu}. For specific models such as $\nu$MSM, the production constraint is typically more stringent than the BBN bound~(a higher lower bound on the mixing angle).  For generality and to be conservative, we only consider the BBN bound. 

\par Sterile neutrino DM produced via mixing can also be a warm DM candidate. While warm DM could be a potential solution to address some small-scale problems of cold DM cosmology~\cite{Horiuchi:2015qri, Bozek:2015bdo, Lovell:2016nkp}, astrophysical observations such as satellite counts or Ly-$\alpha$ forests also constrain DM from being too warm~\cite{Schneider:2016uqi, Baur:2017stq, Cherry:2017dwu}, which puts comparable mixing angle-dependent lower limits on the DM mass that are stronger than the more robust model-independent phase-space constraints~\cite{Tremaine:1979we, Horiuchi:2013noa}. We thus only consider the the satellite-counts constraints from Ref.~\cite{Cherry:2017dwu}.

Figure~\ref{fig:sndm_constraint} also shows the production and warm DM constraints. 
In the context of sterile neutrino DM produced via neutrino mixing, our M31 X-ray constraint reduces the previously available parameter space by close to one-third. 

\section{\label{sec:conclusions} Conclusions and Outlook}
We search for X-ray lines from sterile neutrino DM decay using 1.2\,Ms of combined \nustar{} M31 observations. We consistently include the focused~(2-bounce) FOV, which enhances the sensitivity compared to previous works that considered only the unfocused~(0-bounce) FOV. We also opt to statistically combine the sensitivities of individual observations. Compared to a usual stacking analysis, this reduces the potential systematic error from stacking spectra with different underlying continuum spectra. 

We see no evidence of DM signals, and thus report upper limits for photon line energies 5--100\,keV, or sterile neutrino DM masses 10--200\,keV.  Specifically, the constraints are improved the most in the mass range $\sim 12$--$20$\,keV. 
For more general DM candidates, we also report limits in decay rate and annihilation cross section. For sterile neutrino DM produced via mixing, we reduce the available parameter space by close to one-third. 

We have demonstrated that adding the 2-bounce component can meaningfully enhance the DM sensitivity, up to a factor of 1.4.  However, in this work the enhancement is not optimal, as the center part of the 2-bounce FOV is removed due to the presence of point sources.  In the future, for other observations where such cuts are not needed, the 2-bounce contribution can be increased by a factor $\sim 2$. The 2-bounce contribution could also be higher if the FOV is pointed at regions with more concentrated DM distribution.

Our statistical analysis approach in this work is mainly aimed to reduce systematic errors from the stacking process.  However, in principle, it can also be used to combine completely different observations, such as Galactic Center and M31, after they are properly modeled. This will further enhance the DM search sensitivity.    

In the future, especially with an improved understanding of the low-energy \nustar{} instrumental background model, we anticipate \nustar{} will be a powerful tool to test the remaining parameter space window shown in Fig.~\ref{fig:sndm_constraint}, and perhaps also test the tentative 3.5-keV line signal. Ruling out this window would mean that sterile neutrino DM cannot be simply produced via neutrino mixing.  However, we stress that this would not rule out generic sterile neutrinos as a DM candidate. Nevertheless, this would rule out the $\nu$MSM model~\cite{Asaka:2005an, Asaka:2006nq, Canetti:2012vf,Canetti:2012kh}, which was proposed to simultaneously explain the nature of DM, baryongenesis, and the origin of neutrino mass. 

\section*{\label{sec:acknowledgements} Acknowledgments}

\par The computational aspects of this work made extensive use of the following packages: \texttt{SAOImage DS9} distributed by the Smithsonian Astrophysical Observatory; the \texttt{SciPy} ecosystem \cite{scipy}, particularly \texttt{Matplotlib} and \texttt{NumPy}; and \texttt{Astropy}, a community-developed core Python package for Astronomy \citep{astropy:2013, astropy:2018}. This research has made use of data and/or software provided by the High Energy Astrophysics Science Archive Research Center (HEASARC), which is a service of the Astrophysics Science Division at NASA/GSFC and the High Energy Astrophysics Division of the Smithsonian Astrophysical Observatory. Portions of this work include observations obtained with the \nustar{} mission, a project led by the California Institute of Technology,
managed by JPL, and funded by NASA, as well as \xmm{}, an ESA science mission with instruments and contributions directly funded by ESA Member States and NASA.

\par We thank Shuo Zhang for helpful comments and discussions. K.C.Y.N.\ is supported by a Croucher Fellowship and a Benoziyo Fellowship. B.M.R.\ receives support from MIT Department of Physics and Dean of Science Fellowships. K.P.\ receives support from the Alfred P.\ Sloan Founation. J.F.B.\ is supported by NSF Grant No.\ PHY-1714479. S.H.\ is supported by the U.S.\ Department of Energy under Award No.\ DE-SC0018327.  R.K. acknowledges support from the Russian Science Foundation~(Grant No. 19-12-00396).

\appendix

\section{Stacked analysis}\label{sec:stacked}

\par As described in Sec.~\ref{sec:constraints}, to derive a constraint on a potential DM line flux, we fit each observation separately, and statistically combine the flux constraints for each observation to produce the results shown in Fig.~\ref{fig:sndm_constraint}. This approach allows us to model the background for each observation separately, reducing the systematic effects of combining backgrounds, which may vary for each observation. As a cross-check, we also follow our previous approach in analyzing Galactic Center data \cite{Perez:2016tcq}---\textit{i.e.}, stacking the individual spectra from each observation and deriving DM limits from the stacked spectra. In this section, we discuss the results derived from stacking the spectra, and show that they are consistent with the statistical-combination approach used in Sec.~\ref{sec:constraints}.

\par We use the \texttt{addascaspec} tool to stack the spectra for FPMA/B separately, as each FPM has slightly different instrument responses and internal backgrounds. The 0-bounce and 2-bounce effective areas~(cm$^2$) and solid angles~(deg$^2$) for the stacked spectra are taken to be the exposure-time-weighted averages of the values from the individual observations. The best-fit reduced $\chi^2$ for the stacked spectra, in the null DM hypothesis, is $\sim 2.6$ for each FPM (239 d.o.f.~each, for 200 logarithmically-spaced energy bins per decade), significantly worse than that of the individual observations. As the statistical errors are small ($\lesssim 5\%$), systematic effects dominate. In particular, we find large positive residuals for $E< 6$ keV and $\sim 15$--$30$ keV, similar to those identified in Sec.~\ref{sec:combined_analysis} using the statistical-combination approach. This reinforces the point that the instrumental background model is not sufficient to describe these regions, and the fit cannot be improved via better analysis procedures. Additionally, we find that the derived 2-bounce M31 flux, relative to the (fixed) CXB components, is consistent between the statistical-combination and stacked approaches, as is the derived 2-bounce M31 power-law index. We then use the same line-search procedure on the stacked spectra as was described in Sec.~\ref{sec:combined_analysis} to derive 95\% one-sided upper limits on any DM flux (in the case of a null detection) or to find potential DM signals. We find line-like signals in the same energy ranges as were obtained using the statistical-combination method, though as was discussed previously, a DM interpretation for these features is implausible.

\par Overall, we find that our statistical-combination and stacking approaches provide constraints in the mass-mixing angle plane that are consistent at the $\sim$10\% level, as shown in Fig.~\ref{fig:gamma}. This supports our use of the statistical-combination procedure to derive DM limits, as it provides a better fit (\textit{i.e.}, improved $\chi^2$/d.o.f.) to the data without the inclusion of an overall systematic uncertainty, which weakened our previous limit \cite{Perez:2016tcq}. Additionally, the statistical-combination procedure will allow us to combine constraints from different regions on the sky in future analyses, as we can independently model the backgrounds in each region.

\section{Constraints on DM Annihilation}\label{sec:annihilation}

In this section, we derive upper limits on the DM annihilation cross section.  While thermally produced $s$-wave weakly interacting massive particles~(WIMPs) that couple to the visible sector are strongly constrained at the keV scale~(see discussion in Ref.~\cite{Leane:2018kjk}), DM with non-standard thermal history could still produce X-ray lines through annihilation~~(\textit{e.g.}, see Refs.~\cite{Brdar:2017wgy, Namjoo:2018oyn}). To find the limit on the annihilation cross section, we consider the case of $\chi\chi\rightarrow \gamma \gamma$, and write the analog of Eq.~(\ref{eq:intensity}), 
\begin{equation}\label{eq:annihilation}
    {\cal I} = \frac{\sigma v}{8 \pi m_{\chi}^{2}}\int d\ell \left(\rho_{\rm MW}^{2} + \rho_{\rm M31}^{2}\right) \frac{dN}{dE} \, ,
\end{equation}
where $\sigma v$ is taken to be the velocity-independent annihilation cross section, and the spectrum is $dN/dE = 2\delta(E-m_{\chi})$.  Then it is straightforward to compute the annihilation version of the J-factors and repeat the analysis to obtain the upper limit. We conservatively neglect the potential J-factor enhancement due to DM substructures~(see, \textit{e.g.}, Refs~\cite{Ng:2013xha, Hiroshima:2018kfv}).

\begin{figure}[t]
    \includegraphics[width = \columnwidth]{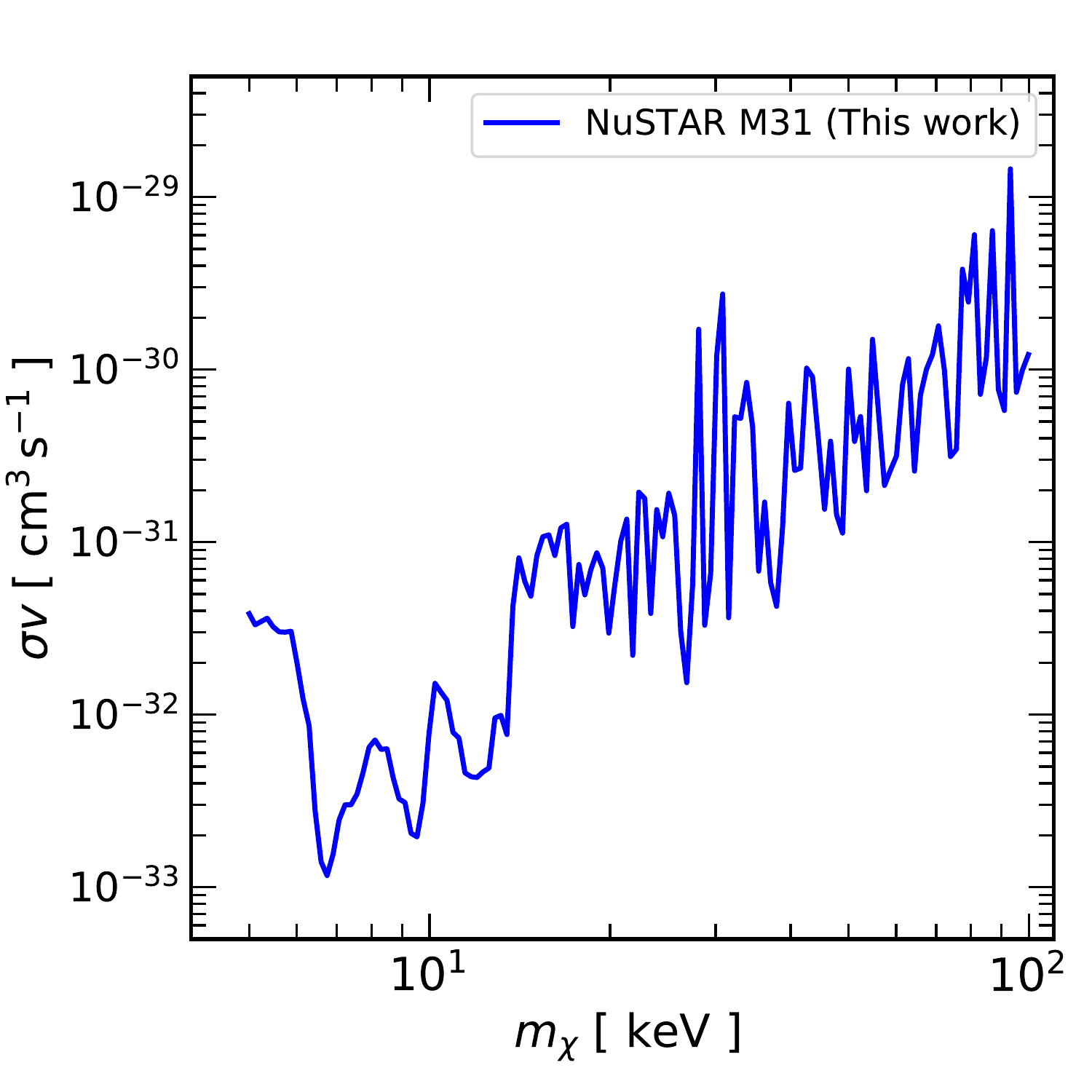}
    \caption{Upper limit on DM annihilation cross sections from combined \nustar{} M31 observations. We have assumed each DM annihilation produces two monoenergetic photons with energy equal to the DM mass. }
    \label{fig:sigmav}
\end{figure}

Figure~\ref{fig:sigmav} shows our derived upper limit on the annihilation cross section.  While we have made several assumptions in Eq.~(\ref{eq:annihilation}), it should be straightforward to translate this limit to other more specific scenarios. 

\bibliography{bib.bib}

\end{document}